\documentclass[a4paper,12pt]{article}
\usepackage{epsfig}
\usepackage{citesort}
\date{\today}
 \normalsize

\def\be{\begin{equation}}
\def\ee{\end{equation}}
\def\bea{\begin{eqnarray}}
\def\eea{\end{eqnarray}}

\def\lsim{\raise0.3ex\hbox{$\;<$\kern-0.75em\raise-1.1ex\hbox{$\sim\;$}}}
\def\gsim{\raise0.3ex\hbox{$\;>$\kern-0.75em\raise-1.1ex\hbox{$\sim\;$}}}

\def\ie{{\it i.e.}}
\begin{document}
\renewcommand{\thefootnote}{\fnsymbol{footnote}}
\rightline{\today} \vspace{.3cm} {\large
\begin{center}
{\large \bf Low scale $B-L$ extension of the Standard Model at the
LHC}
\end{center}}
\vspace{.3cm}
\begin{center}
Shaaban Khalil\\
\vspace{.3cm} {\small \it Center for Theoretical Physics, British
University in Egypt, Cairo, 11837, Egypt.}\\
{\small \it Department of Mathematics, Faculty of Science, Ain
Shams
University,  Cairo, 11566, Egypt.}\\
\end{center}
\vspace{.3cm} \hrule \vskip 0.3cm
\begin{center}
\small{\bf Abstract}\\[3mm]

\begin{minipage}[h]{14.25cm}
The fact that neutrinos are massive indicates that the Standard
Model (SM) requires extension. We propose a low energy ($\lsim
TeV$) $B-L$ extension of the SM, which is based on the gauge group
$SU(3)_C\times SU(2)_L \times U(1)_Y \times U(1)_{B-L}$. We show
that this model provides a natural explanation for the presence of
three right-handed neutrinos in addition to an extra gauge boson
and a new scalar Higgs. Therefore, it can lead to very interesting
phenomenological implications different from the SM results which
can be tested at the LHC. Also we analyze the muon anomalous
magnetic moment in this class of models. We show that one-loop
with exchange $Z^{\prime}$ may give dominant new contribution $
\sim few~ \times 10^{-11}$.
\end{minipage}
\end{center}
\vskip 0.3cm \hrule \vskip 0.5cm
\section{{\large \bf Introduction}}
There are at present two pieces of evidence which hint at physics
beyond the SM: $(i)$ The solid evidence for neutrino oscillations,
pointing towards non-vanishing neutrino masses. In the SM,
neutrinos are massless due to the absence of right-handed
neutrinos and the exact $B-L$ conservation. $(ii)$ The strength of
CP violation in the SM is not sufficient to generate the observed
baryon asymmetry in the universe. The neutrino mass puzzle and the
baryon asymmetry problem can be readily solved by introducing
right-handed neutrinos. Right-handed neutrinos lead to the see-saw
mechanism. This mechanism explains, in an elegant way, why neutrinos are
much lighter than the other elementary fermions. Also, the new
complex phases in the leptonic sector can generate lepton
asymmetry, which is converted to baryon asymmetry,
 through the decay of the right-handed neutrinos.

The tremendous success of gauge symmetry in describing the SM
makes us believe that any extension of this model should be
through the extension of its gauge symmetry. The minimal type of
this extension is based on the gauge group $G_{B-L}\equiv
SU(3)_C\times SU(2)_L\times U(1)_Y\times U(1)_{B-L}$. In fact, the
SM is characterized by possessing, at the renormalizable level, a
global $U(1)_{B-L}$ symmetry. If this symmetry is gauged, the
existence of three $SM$ singlet fermions (we call them right
handed neutrinos) are predicted by the anomaly cancellation
conditions, as crucial ingredients for the consistency of the
model. In addition, the model also contains an extra gauge boson
corresponding to $B-L$ gauge symmetry and an extra SM singlet
scalar (heavy Higgs). This may change significantly the SM
phenomenology and lead to interesting signatures at the LHC.

In fact, it is very difficult to imagine that the new physics
beyond the SM that nature adopted for generating neutrino masses
will not be manifested in any other low energy process. Indeed,
this is the case when three right-handed neutrinos are randomly
added to the SM spectrum. However, with this minimal extension of
the SM gauge group, one finds that the neutrino masses are
strongly related to some other low energy processes. It is worth
mentioning that the extra gauge and Higgs bosons contained in this
class of models are playing important role in establishing this
relation between the mechanism of generating the neutrino masses
and low energy physics consequences.

In this letter we reappraise the low scale scenario of $B-L$
extension of the SM. We will show that this class of model can
account for the experimental results of the light neutrino masses
and their large mixing. The TeV scale $B-L$ symmetry breaking and
see-saw mechanism have not been considered in much details in the
literature. There were some attempts in the past for analyzing the
$B-L$ extension of the SM \cite{B-L}. However, these attempts were
focused on the neutrino masses, based on old experimental results,
in minimal or non-minimal extensions of the SM that include $B-L$
symmetry. Here, we perform a complete analysis for the low scale
($\lsim 1 TeV$) $B-L$ minimal extension of the SM, where the
effects of the new spectrum in different sectors are
simultaneously taken into account.


The paper is organized as follows. In section 2 we discuss the
$B-L$ extension of the SM with especial emphasis for the
simultaneous breaking of $U(1)_{B-L}$ and $SU(2)_L\times U(1)_Y$.
The analysis of the Higgs bosons in this model is given in section
3. Section 4 is devoted for the neutrino masses and mixing in this
low scale $B-L$ extension of the SM. In section 5 we discuss the
new gauge boson $Z^{\prime}$ in this model and its implication on
the muon anomalous magnetic moment. Finally, we give our
conclusions in section 6.

\section{{\large \bf$B-L$ extension of the SM}}

In this section we discuss the particle content and the
spontaneous symmetry breaking in the minimal extension of the SM
based on the gauge group $G_{B-L} \equiv SU(3)_C\times
SU(2)_L\times U(1)_Y\times U(1)_{B-L}$. The invariance of the
lagrangian under this gauge symmetry implies the existence of a
new gauge boson (beyond the SM ones) $C_\mu$. Also in order to
ensure that $U(1)_{B-L}$ is anomaly free, three SM singlet
fermions must be introduced. These singlet fermions are usually
called right-handed neutrinos and denoted by $\nu_{R_i}$. In this
respect, the global ($B-L$) symmetry in the SM is gauged and the
scale of breaking of this symmetry provides a natural
identification to what is called seesaw scale.

The Lagrangian of the leptonic sector in the minimal extension of the SM $G_{B-L}$ is given by%
\bea%
{\cal L}_{B-L}&=&-\frac{1}{4} C_{\mu\nu}C^{\mu\nu} + i~ \bar{l}
D_{\mu} \gamma^{\mu} l + i~ \bar{e}_R D_{\mu} \gamma^{\mu} e_R +
i~ \bar{\nu}_R D_{\mu} \gamma^{\mu} \nu_R + (D^{\mu}
\phi)(D_{\mu} \phi) \nonumber\\
&+& (D^{\mu} \chi)(D_{\mu} \chi)- V(\phi, \chi)- \Big(\lambda_e
\bar{l} \phi e_R + \lambda_{\nu} \bar{l} \tilde{\phi} {\nu}_R +
\frac{1}{2} \lambda_{\nu_R} \bar{\nu^c}_R \chi \nu_R + h.c.\Big),%
\label{lagrangian}
\eea %
where $C_{\mu\nu} = \partial_{\mu} C_{\nu} - \partial_{\nu}
C_{\mu}$ is the field strength of the $U(1)_{B-L}$. The covariant
derivative $D_{\mu}$ is generalized by adding the term $i g^{''}
Y_{B-L} C_{\mu}$, where $g^{''}$ is the $U(1)_{B-L}$ gauge
coupling constant and $Y_{B-L}$ is the $B-L$ quantum numbers of
involved particles. The $Y_{B-L}$ for leptons and Higgs are given
by: $Y_{B-L}(l) =-1$, $Y_{B-L}(e_R) =-1$, $Y_{B-L}(\nu_R) =-1$,
$Y_{B-L}(\phi) =0$ and $Y_{B-L}(\chi) =2$. In
Eq.(\ref{lagrangian}), $\lambda_{e}$, $\lambda_{\nu}$ and
$\lambda_{\nu_R}$ refer to $3\times 3$ Yakawa matrices.


The Higgs sector of this model should contain one $SU(2)_L$
singlet complex scaler field $\chi \equiv (1,1,0,2)$ that can
spontaneously break the $U(1)_{B-L}$ symmetry and one $SU(2)_L$
doublet $\phi \equiv (1,2,1,0)$ to break the $SU(2)_L \times
U(1)_Y$ symmetry down to $U(1)_{em}$. The non-vanishing vacuum
expectation value (vev) of $\chi$: $|\langle\chi\rangle|= v'/\sqrt
2$ is assumed to be larger than the vev of the Higgs field $\phi$:
$|\langle\phi^0\rangle|= v/\sqrt 2$. It is remarkable that the
scale of $U(1)_{B-L}$ symmetry breaking, $v'$, is not fixed.
However, as we will see, the mass of the $U(1)_{B-L}$ gauge boson
$C_\mu$ is given in terms of $v'$. Therefore it can be bounded
from below by the experimental search for extra neutral gauge
boson.

In order to analyze the $B-L$ and electroweak symmetry breaking,
we consider the most general Higgs potential invariant under
these symmetries, which is given by%
\bea%
V(\phi,\chi)&=&m_1^2 \phi^\dagger \phi+m_2^2 \chi^\dagger
\chi+\lambda_1 (\phi^\dagger\phi)^2+\lambda_2
(\chi^\dagger\chi)^2\nonumber\\
&+&\lambda_3
(\chi^\dagger\chi)(\phi^\dagger\phi),%
\eea%
where $\lambda_3 > - 2 \sqrt{\lambda_1\lambda_2}$ and $\lambda_1,
\lambda_2 \geq 0$, so that the potential is bounded from below.
This is the stability condition of the potential. Furthermore, in
order to avoid that $\langle \phi \rangle=\langle \chi \rangle=0$
be a local minimum, we have to require that $\lambda_3^2 < 4
\lambda_1 \lambda_2$. As in the usual Higgs mechanism of the SM,
the vevs $v$ and $v'$ can not be emerged unless negative squared
masses, $m^2_{1} < 0$ and $m^2_{2} <0$, are assumed.
In this case, the non-zero minimum is given by %
\bea %
v^2= \frac{4 \lambda_2 m_1^2 - 2 \lambda_3 m_2^2}{\lambda_3^2
-4 \lambda_1\lambda_2},~~~~~~~ v'^2 = \frac{-2 (m_1^2 +\lambda_1 v^2)}{\lambda_3}. %
\eea %

%
\begin{figure}[t]
\begin{center}
\epsfig{file=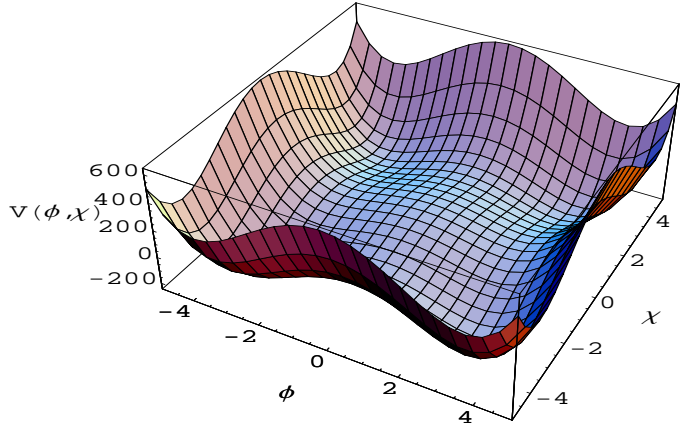, width=9cm, height=7cm, angle=0}
\end{center}
\vskip -0.75cm \caption{The $B-L$ and electroweak symmetry
breaking minima with $m_1^2 < 0$, $m_2^2 <0$ and $\lambda_3 \sim
-0.5 \sqrt{\lambda_1 \lambda_2}$.} \label{vev}
\end{figure}

As can be seen from the above expressions, the vevs $v$ and $v'$
can not be emerged unless negative squared masses, $m^2_{1} < 0$
and $m^2_{2} <0$, are assumed, as in the usual Higgs mechanism of
the SM. It is also interesting to note that with $\lambda_3>0$, it
is possible to generate non-vanishing vev $v'$, \ie, the $B-L$
symmetry is broken, while the vev $v=0$. This type of minimum
corresponds to the scenario of two stages symmetry breaking at
different scale, with $v' \gg v$. In our analysis, we are
interested in the case of low scale $v'$, which might be of order
the electroweak. Therefore, we will focus on the following region
of mixing coupling $\lambda_3$: $0 > \lambda_3 > -2
\sqrt{\lambda_1\lambda_2}$. In Fig1. we present the vevs of the
potential $V(\phi,\chi)$ in case of $m_{1,2}^2 < 0$, $\lambda_3
\simeq -0.5 \sqrt{\lambda_1 \lambda_2}$ and $\lambda_{1,2} \sim
{\cal O}(1)$

After the $B-L$ gauge symmetry breaking, the gauge field $C_{\mu}$
(will be called $Z^{\prime}$ in the rest of the paper) acquires
the
following mass: %
\be%
M_{Z^{\prime}}^2=4 g''^2 v'^2 .%
\ee%
The high energy experimental searches for an extra neutral gauge
boson impose lower bounds on this mass. The CDF limit
\cite{Abe:1997fd} leads to $M_{Z^{\prime}} \gsim O(600-800)$ GeV.
However, since LEP II was $e^+ e^-$ collider, it constrain
strongly the extra-gauge boson that coupled significantly with
electrons. Therefore, LEP II provides the most stringent
constraint on $B-L$ gauge boson and implies that
\cite{Carena:2004xs}\footnote{We would like to thank
A. Dobrescu for drawing our attention to this limit.}) %
\be%
M_{Z^{\prime}} /g'' > 6 ~ \rm{TeV}. %
\ee%
This implies that $v' \gsim O(\rm{TeV})$. Moreover, if the
coupling $g^{''} < O(1)$, one can still obtain $m_{Z^{\prime}}
\gsim O(600)$ GeV.
\section{{\large \bf Higgs in $B-L$ extension of the SM}}

Now we turn to the Higgs sector in this class of models. As
mentioned, the Higgs scalar fields in $G_{B-L}$ consists of one
complex $SU(2)_L$ doublet and one complex scalar single, \ie, six
scalar degrees of freedom. When the $B-L$ and electroweak
symmetries are broken, four of them are eaten by $Z^{\prime}$,
$Z^0$ and $W^{\pm}$ bosons and two scalar bosons ($\phi,\chi$)
remain as physical degrees of freedom. From the mass terms in the
scalar potential, one finds
the following mass matrix for $\phi$ and $\chi$: %
\be%
\frac{1}{2} M^2 (\phi,\chi) =\left(%
\begin{array}{cc}
  \lambda_1 v^2 & \frac{\lambda_3}{2} v v' \\
 \frac{\lambda_3}{2} v v' & \lambda_2 v'^2 \\
\end{array}%
\right). %
\ee %
Therefore, the mass eigenstates fields $H$ and $H^{\prime}$ are
given as %
\be %
\left(\begin{array}{c} H\\
H^{\prime} \end{array} \right)= \left(\begin{array}{cc} \cos \theta & - \sin\theta\\
\sin\theta & \cos \theta \end{array} \right)  \left(\begin{array}{c} \phi\\
\chi \end{array} \right),%
\ee where the mixing angle $\theta$ is defined by %
\be %
\tan 2 \theta = \frac{\vert \lambda_3 \vert v v'}{\lambda_1 v^2 -
\lambda_2 v'^2}. %
\ee %
The masses of $H$ and $H^{\prime}$ are given by
\be%
m^2_{H,H^{\prime}} = \lambda_1 v^2 + \lambda_2 v'^2 \mp
\sqrt{(\lambda_1 v^2 - \lambda_2 v'^2)^2 +\lambda_3^2 v^2 v'^2}.%
\ee%
We call $H$ and $H^{\prime}$ light and heavy Higgs bosons
respectively. From these expressions, it is clear that $\lambda_3$
is the measuring of the mixing between the SM Higgs and the $B-L$
extra Higgs. For instance, with $\lambda_3=0$, there is no mixing
and the Higgs masses is given by $m_{\phi} =\sqrt{2 \lambda_1} v$,
as in the SM, and $m_{\chi} = \sqrt{2 \lambda_2} v'$. While for
$\lambda_3 \neq 0$, one finds that the light Higgs mass becomes
smaller than the SM prediction.

Due to the mixing between the two Higgs bosons, the usual SM
couplings among the SM-like Higgs $H$ and the SM fermions and
gauge bosons are modified as follows: %
\bea %
g_{H ff} &=&  i \frac{m_f}{v}
\cos \theta , \nonumber\\
g_{H VV} &=& - 2 i  \frac{M_V^2}{v} \cos \theta , \nonumber\\
g_{H H VV} &=& - 2 i \frac{M^2_V}{v^2} \cos^2 \theta .%
\eea %
Here, we follow the conventions of Ref. \cite{Djouadi:2005gi} for
the Feynman rules of the vertices. In addition, there are new
couplings among the extra Higgs, $H^{\prime}$,
and the SM particles :%
\bea %
g_{H^{\prime} ff} &=&  i \frac{m_f}{v} \sin \theta , \nonumber\\
g_{H^{\prime} VV} &=& - 2 i \frac{M_V^2}{v} \sin \theta , \\
g_{H H^{\prime} VV} &=& - 2 i \frac{M^2_V}{v^2} \cos \theta
\sin\theta. \nonumber%
\eea %
We have adopted the first order approximation of the mixing angle
$\theta$. Therefore, terms of order $\sin^n\theta$, $n \geq 2$
have been neglected. Furthermore, the right-handed neutrino and
the $B-L$ gauge boson $Z^{\prime}$ are now coupled
with both $H$ and $H^{\prime}$ as follows:%
\bea %
g_{H \nu_R \nu_R} &=& - i \frac{m_{\nu_R}}{v'} \sin \theta, ~~~~~
~~~~~ ~~~~~ g_{H^{\prime} \nu_R \nu_R} = i \frac{m_{\nu_R}}{v'}
\cos
\theta, \nonumber\\
g_{H^{\prime} Z^{\prime} Z^{\prime}} &=& - 2 i
\frac{M_{Z^{\prime}}}{v'} \cos \theta ,~~~~~ ~~~~~ ~~~~g_{H
Z^{\prime}Z^{\prime}}=  2 i \frac{M_{Z^{\prime}}}{v'} \sin \theta,\\
g_{H^{\prime} H^{\prime} Z^{\prime}Z^{\prime}} &=& -2 i
\frac{M^2_{Z^{\prime}}}{v'^2} \cos^2\theta ~~~~ ~~~~~ ~~~~~ g_{H
H^{\prime} Z^{\prime} Z^{\prime}} =2 i
\frac{M^2_{Z^{\prime}}}{v'^2} \cos\theta \sin\theta.\nonumber
\eea %
Finally we consider the Higgs self interaction vertices. One can
easily prove that they are given by%
\bea %
g_{H^3} &=& 6 i \left(\lambda_1 v \cos^3\theta
-\frac{\lambda_3}{2} v'\cos^2\theta \sin\theta \right),\nonumber\\
g_{H'^3} &=& 6 i \left(\lambda_2 v' \cos^3\theta +
\frac{\lambda_3}{2} v\cos^2\theta \sin\theta \right),\nonumber\\
g_{H^4} &=& 6 i \lambda_1 \cos^4\theta,~~~~~~ ~~~~~
g_{H'^4} = 6 i \lambda_2 \cos^4\theta,\nonumber\\
g_{H H'^2} &=& 2 i \left( \frac{\lambda_3}{2} v \cos^3\theta +
\lambda_3 v' \cos^2\theta \sin\theta - 3 \lambda_2
v' \cos^2\theta \sin\theta \right),\nonumber\\
g_{H^2 H'} &=& 2 i \left( \frac{\lambda_3}{2} v' \cos^3\theta -
\lambda_3 v \cos^2\theta \sin\theta + 3
\lambda_1 v \cos^2\theta \sin\theta \right),\nonumber\\
g_{H^2H'^2} &=&  i \lambda_3 \cos^4\theta.
\eea %

These new couplings lead to a different Higgs phenomenology from
the well known one predicted by the SM. In Ref.\cite{Emam:2007dy},
it is shown that in this class of models the cross sections of the
SM-like Higgs production are reduced by $~ 20\% - 30\%$ in the
mass range of $~ 120 - 250$ GeV compared to the SM results. While,
the implications of the $B - L$ extension to the SM do not change
the decay branching ratios. Moreover, the extra Higgs has
relatively small cross sections, but it is accessible at LHC.

On the other hand, there are now two fine-tuning problems
associated to the electroweak and $B-L$ symmetry breaking. These
problems are based on the sensitivity of the Higgs boson masses
$m_{H}$ and $m_{H^{\prime}}$ to quadratic divergences. It is well
known that within the SM, the one loop radiative correction to the Higgs leads to  %
\be %
\Delta M_H^2 \propto \frac{3 \Lambda^2}{8 \pi^2 v^2} \left[M_H^2 +
2 M_W^2 + M_Z^2 - 4 m_t^2 \right].%
\ee %
Thus, in order to avoid a fine-tuning between this correction and
the tree level value of$M_{H}^2$, an upper bound $\Lambda$ is
obtained. For example for $M_{H} \simeq 115 -200$ GeV one finds %
\be %
\left\vert \frac{\Delta M_H^2}{M_H^2} \right\vert \leq 10
~~~~~~~~~~ \rightarrow ~~~~~~~~~~~\Lambda_{{\rm SM}} \lsim 2- 3~ {\rm TeV}.%
\label{SMcut}
\ee %
In addition, the cancellation of the quadratic divergences may
lead to a prediction for the Higgs mass as suggested by Veltman
\cite{Veltman:1980mj}. Indeed with $M_{H} \sim (320~ GeV)$ the
above correction would vanish. However, this conclusion valid only
at the one loop level and one has to ensure that it can also be
valid at higher orders.

\begin{figure}[t]
\begin{center}
\epsfig{file=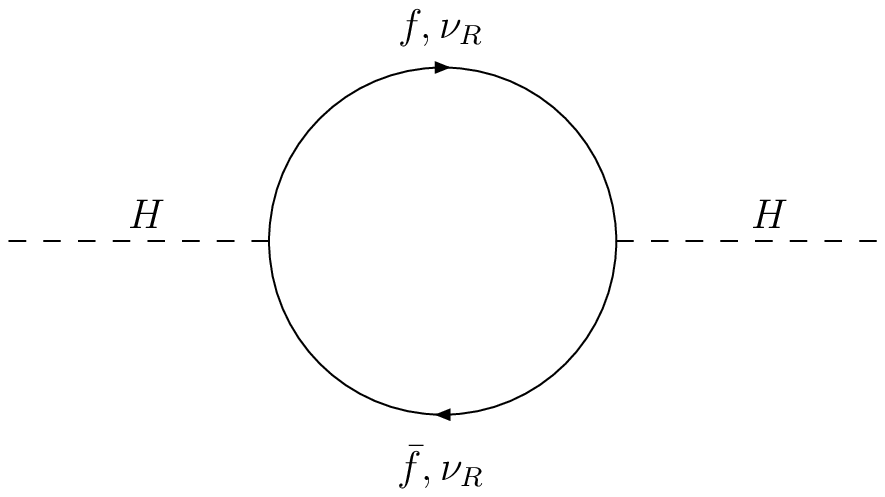, width=5cm, height=2.5cm, angle=0}~
\epsfig{file=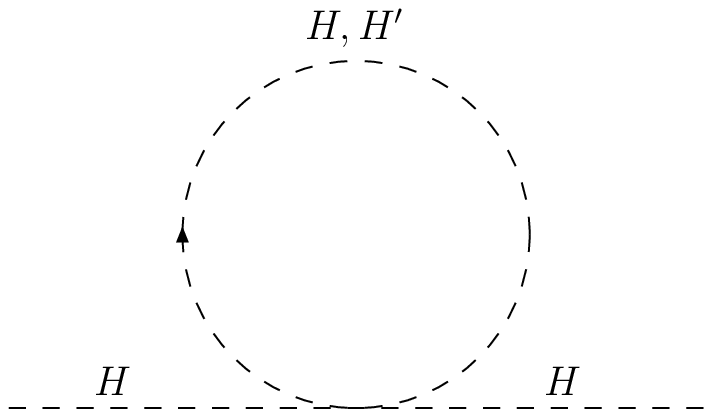, width=4cm, height=2.2cm, angle=0}~
\epsfig{file=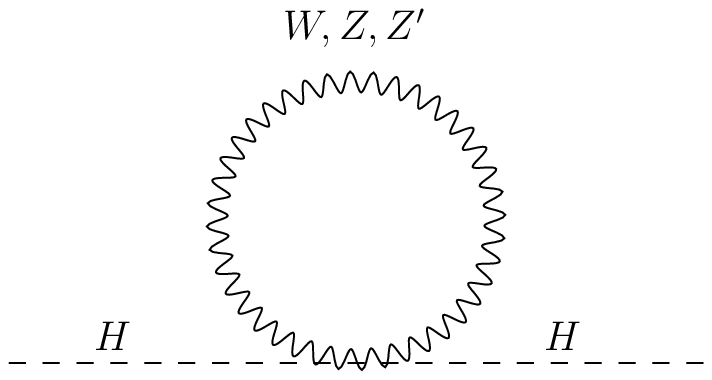, width=4cm, height=2.2cm, angle=0}
\end{center}
\caption{Feynman diagrams that lead to one-loop quadratic
divergences for the light Higgs boson mass in $B-L$ extension of
the SM.} \label{DMSM}
\end{figure}

In the $B-L$ extension of the SM, the Feynman diagrams that lead
to one-loop quadratic divergences for the light Higgs boson mass
are given in Fig. 1, where Higgs boson, gauge boson and fermion
are running in the loop. It is worth noting that three new
diagrams different from the SM ones are now involved. These
diagrams include loops with right-handed neutrino, $Z^{\prime}$
gauge boson and heavy
Higgs boson. In this case, one obtains%
\bea %
\Delta M_{H}^2 &\propto& \frac{3 \Lambda^2}{8 \pi^2} \left[\left(2
\lambda_1 + \frac{\lambda_3}{3}\right) \cos^4\theta +
\left(\frac{2 M_W^2+ M_Z^2}{v^2}\right) \cos^2\theta
\right.\nonumber\\
&+& \left.\frac{M_{Z^{\prime}}^2}{v'^2} \sin^2 \theta- 4
\left(\frac{m_t^2}{v^2} \cos^2\theta+
\frac{m_{\nu_R}^2}{v'^2}\sin^2\theta\right)\right].~%
\eea %
Now the condition of no fine tuning (with $m_H \simeq 200~ GeV$ ,
$v'\simeq 1~ TeV$ and $\cos\theta^2 \simeq 0.8$) implies that %
\be %
\Lambda_{{\rm SM}} \approx ~ {\cal O}(10)~ {\rm TeV}. %
\ee %
The exact limit depends on the values of the masses
$M_{Z^{\prime}}$ and $M_{\nu_R}$. In general, in the $B-L$ model,
the SM cut off scale becomes higher than the limit obtained in
Eq.(\ref{SMcut}). This new bound is more consistent with the
experimental lower bounds usually imposed to suppress higher order
operators. Therefore, it is natural to avoid in this class of
model, what is called as a little hierarchy problem. Also, now the
mass of the SM-like Higgs is not predictable from the cancellation
condition of quadratic divergences.
\bea %
\Delta M_{H^{\prime}}^2 &\propto& \frac{3 \Lambda^2}{8 \pi^2}
\left[\left(2 \lambda_2 + \frac{\lambda_3}{3}\right) \cos^4\theta
+ \left(\frac{2 M_W^2+ M_Z^2}{v^2}\right) \sin^2\theta
\right.\nonumber\\
&+& \left.\frac{M_{Z^{\prime}}^2}{v'^2} \cos^2 \theta- 4
\left(\frac{m_t^2}{v^2} \sin^2\theta+
\frac{m_{\nu_R}^2}{v'^2}\cos^2\theta\right)\right].~%
\eea %
It is interesting to note that to avoid a fine-tuning between
$M^2_{H^{\prime}}$ and $\Delta M_{H^{\prime}}^2$, one finds, for
$M_{H^{\prime}} \simeq 1$ TeV, the following bound %
\be %
\left\vert \frac{\Delta M_{H^{\prime}}^2}{M^2_{H^{\prime}}}
\right\vert \leq 10
~~~~~~~~~~ \rightarrow ~~~~~~~~~~~\Lambda_{B-L} \lsim 30~ {\rm TeV}.%
\label{SMcut}
\ee %
Moreover, Veltman cancellation condition for the extra Higgs
quadratic divergences leads to (in the limit of very small mixing)%
\be %
M_{H^{\prime}}^2 \approx 4 M_{\nu_R}^2 - M_{Z^{\prime}}^2. %
\ee %
Thus, one may conclude that $M_{H^{\prime}} \simeq {\cal O}(1)$
TeV.

\section{{\large \bf Neutrino masses and mixing in low scale $B-L$ extension of the
SM}}

In this section we analyze the neutrino masses and mixing in the
low scale gauge $B-L$ extension of the SM. In this class of
models, the neutrino masses may be generated through a TeV scale
seesaw mechanism.

After $U(1)_{B-L}$ symmetry breaking, the Yukawa interaction term
in Eq.(\ref{lagrangian}): $\lambda_{\nu_R}\chi\nu_{R}\nu_{R}$
leads, as usual, to right handed neutrino mass: $ M_R =
\frac{1}{\sqrt 2}\lambda_{\nu_R} v' $. Also the electroweak
symmetry breaking implies Dirac neutrino mass term :
$m_D=\frac{1}{\sqrt{2}}\lambda_\nu v$. Therefore, the mass matrix
of the left and right-handed neutrino is
given by %
\be
\left(%
\begin{array}{cc}
  0 & m_D \\
  m_D & M_R \\
\end{array}%
\right). %
\ee%

Since $M_R$ is proportional to $v'$ and $m_D$ is proportional to
$v$ \ie, $M_R \gg m_D$, the digitalization of the mass matrix
leads to the following mass for the light and heavy neutrinos respectively: %
\bea%
m_{\nu_L} &=& -m_D M_R^{-1} m_D^T ,\\
m_{\nu_H} &=& M_R . %
\eea %
Thus, $B-L$ gauge symmetry can explain the presence of three right
handed neutrinos and provide a natural framework for the seesaw
mechanism. However as mentioned the scale of B-L and hence the
mass $M_R$ of $\nu_R$ remains arbitrary. It is often assumed a
very large scale for $B-L$ symmetry breaking, \ie,
$M_R\sim10^{15}$ GeV in order to explain the atmospheric and solar
neutrino data. It is important to note that such a large scale may
be necessary if the Dirac neutrino masses are assumed to be of
order ${\cal O}(100)$ GeV. However, there is no any low energy
evidence that indicates that the Dirac masses should be of that
order. On the contrary, if one tries to establish a flavor
symmetry between charged and neutral leptons as in quark sector
between up and down, one finds that the Dirac neutrino masses must
be very small, of order ${\cal O}(10^{-4})$ GeV. This implies that
$M_R$ of order TeV is quite acceptable.

In our analysis, we adopt the basis where the charged lepton mass
matrix and the Majorana mass matrix $M_R$ are both diagonal.
Therefore, one can parameterize $M_R$ as follows%
\be%
M_R=M_{R_3}\left(%
\begin{array}{ccc}
  r_1 & 0 & 0 \\
  0 & r_2 & 0 \\
  0 & 0 & 1 \\
\end{array}%
\right), %
\label{MR}
\ee %
where %
\be %
M_{R_3}=\vert \lambda_{\nu_{R_3}} \vert
\frac{v'}{\surd2}%
\ee %
and %
\be %
r_{1,2}=\frac{M_{R_{1,2}}}{M_{R_3}}=
\left\vert \frac{\lambda_{\nu_{R_{1,2}}}}{\lambda_{\nu_{R_3}}}\right\vert .
\ee %
As can be seen from Eq.(\ref{MR}) that even if $v'$ is fixed to be
of order TeV, the absolute value of $M_R$ is still parameterized
by three known parameters. On the other hand, the Dirac mass
matrix (if it is real) is given in terms of $9$ parameters. Since
$U(1)_{B-L}$ can not impose any further constraint to reduce the
number of these parameters, the total number of free parameters
involved in the light neutrino mass matrix are $12$ parameters.
The solar and atmospheric neutrino oscillation experiments have
provided measurements for the neutrino mass-squared differences
and also for the neutrino mixing angles. At the $3 \sigma $ level,
the allowed ranges are \cite{Mohapatra:2006gs}:
\begin{eqnarray}
\Delta m_{12}^2 &=& (7.9 \pm 0.4) \times 10^{-5} \rm{eV}^2 ,\nonumber\\
\vert \Delta m_{32}^2 \vert &=& (2.4+0.3)\times 10^{-3} \rm{eV}^2 ,\nonumber\\
\theta_{12} &=& 33.9^{\circ} \pm 1.6^{\circ},\\
\theta_{23} &=& 45^{\circ},\nonumber\\
\sin^2 \theta_{13} & \leq & 0.048 .\nonumber
 \label{dm2}
\end{eqnarray}
Therefore, the number of the experimental inputs are at most six:
three neutrino masses (assuming possible ansatze like hierarchy or
degenerate) and three mixing angles (if we assume
$\theta_{13}=0$).

One of the interesting parametrization for the Dirac
neutrino mass matrix is given as follows %
\be %
m_D=\sqrt{M_R} R \sqrt{m_\nu^{diag}} U_{MNS}^\dagger , %
\label{mD}
\ee %
where $m_\nu ^{diag}$ is the physical neutrino mass matrix and
$U_{MNS}$ is the lepton mixing matrix. The matrix R is an
arbitrary orthogonal matrix which can be parameterized-in case of
real $m_D$-in terms of three angles. In Eq.(\ref{mD}), the six
unknown parameters are now given in terms of three masses in $M_R$
and the three angles in $R$. In order to fix these angles, one
need a flavor symmetry beyond the gauge symmetry which is
typically flavor blind. Several types of flavor symmetries have
been discussed in the literatures.

\section{{\large \bf Extra $B-L$ gauge bosnon and muon anomalous magnetic moment}}%

As mentioned above, an extra gauge boson corresponding to $B-L$
gauge symmetry is predicted. In fact, there are many models which
contain extra gauge
bosons~\cite{Cvetic:1995zs,Carena:2004xs,Leike:1998wr,Hewett:1989rm}.
These models can be classifies into two categories depending on
whether or not they arise in a GUT scenario. In some of these
models, the $Z^{\prime}$ and the SM $Z$ are not true mass
eigenstates due to mixing. This mixing induces the couplings
between the extra $Z^{\prime }$ boson and the SM fermions.
However, there is a stringent experimental limit on the mixing
parameter.
In our model of $B-L$ extension of the SM, there is no tree-level $%
Z-Z^{\prime }$ mixing. Nevertheless, the extra $B-L$ $Z^{\prime }$
boson and the SM fermions, are coupled through the non-vanishing
$B-L$ quantum numbers. In Ref.\cite{Emam:2007dy}, it was shown
that within $B-L$ extension of the SM the branching ratios of
$Z^{\prime} \to l^+ l^-$ are of order $\sim 20\%$ compared to
$\sim 3\%$ of the SM $BR(Z\to l^+ l^-)$. Hence, searching for
$Z^{\prime}$ is accessible via a clean dilepton signal at LHC.

Here, we consider the impact of the $Z^{\prime}$ and also
$H^{\prime}$ on the muon anomalous magnetic moment ($a_\mu$). The
$a_\mu$ has recently been determined with a very high precision.
From the the measurement of E821 collaboration at the Brookhaven
National Laboratory the average value of $a_\mu$ is given by \cite{Bennett:2006fi}%
\be%
a_\mu^{\rm{exp}}=(116592080\pm60)\times10^{-11} %
\ee%
This value differs from the SM prediction \cite{Hagiwara:2006jt} by $3.4\sigma$: %
\be %
\Delta a_\mu =a_\mu ^{\rm{exp}}-a_\mu^{\rm{SM}}=(276\pm 81)\times 10^{-11} %
\label{amuresult}
\ee%

\begin{figure}[t]
\begin{center}
\epsfig{file=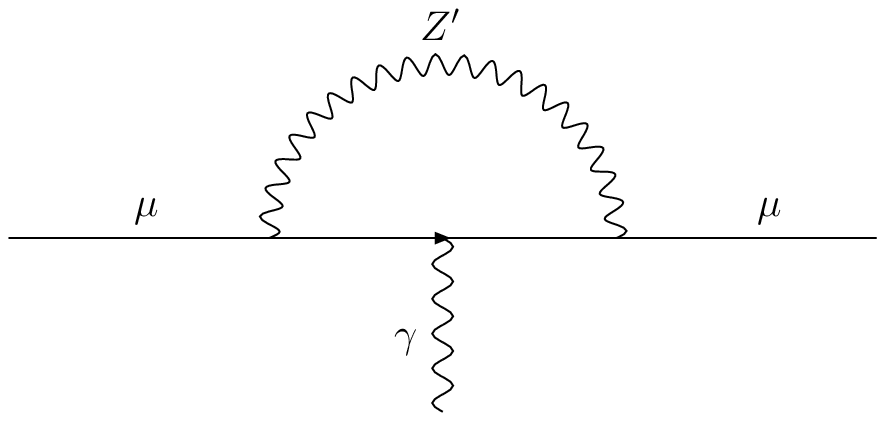, width=7cm, height=3.5cm, angle=0}~~~~~
\epsfig{file=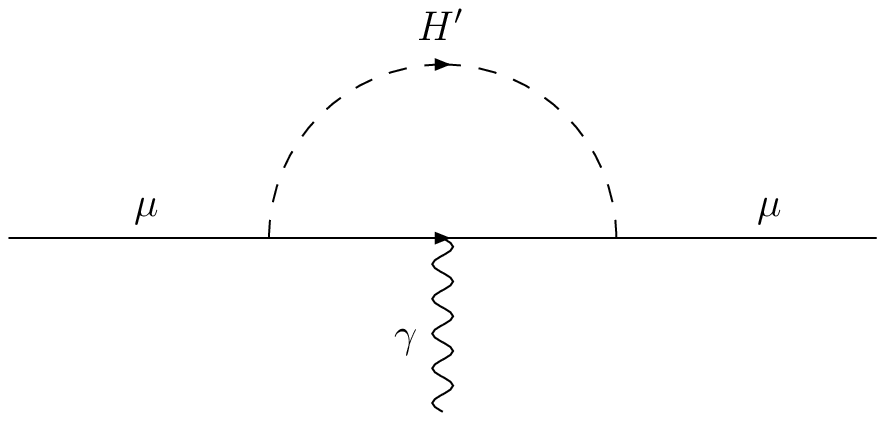, width=7cm, height=3.5cm, angle=0}~
\end{center}
\caption{Feynman diagrams for muon anomalous magnetic moment with
$Z^{\prime}$ and $H^{\prime}$ exchanges.} \label{g-2}
\end{figure}

In $B-L$ extension of the SM, $a_\mu$ can be generated, in
addition to the SM contributions, through other one loop diagrams
mediated by the new gauge boson $Z^{\prime}$ and the extra Higgs
boson $H^{\prime}$ as in Fig. \ref{g-2}. However, the $H^{\prime}$
contribution is proportional to $\lambda_{\mu}^2 \sin\theta^2$,
therefore its contribution is quite small and can be safely
neglected. The $Z^{\prime}$ contribution is given by
\be%
\Delta a_\mu \simeq \frac{g^{''^2}}{12 \pi^2}
\frac{m_{\mu}^2}{M_{Z^{\prime}}^2} \simeq
\frac{m_{\mu}^2}{48 \pi^2 v'^2}%
\ee%
where $m_{\mu}$ is the mass of the muon. For $v'\sim 1$ TeV, one
finds that %
\be %
\Delta a_{\mu}(B-L) \sim 2.1 \times 10^{-11},%
\ee %
which is clearly not enough to account for the deviation between
the experimental result and the SM prediction. Therefore,
confirming this discrepancy in anomalous magnetic moment $a_{\mu}$
would be useful hint for further new physics beyond this minimal
extension of the SM.

\section{Conclusions}

We have analyzed the minimal extension of the SM, which is based
on the $TeV$ scale $B-L$ gauge symmetry. We have shown that this
class of models can give a natural explanation for the $TeV$ scale
seesaw mechanism and very small neutrino masses. We provided a
detail analysis for the simultaneous breaking of the electroweak
and $U(1)_{B-L}$ symmetries. We emphasized that the Higgs sector
of this model is quite rich with possible significant mixing
between the SM-Higgs and the extra-Higgs. This mixing implies
interesting implications, which can be tested at the LHC. Also due
to the fact the $B-L$ gauge boson $Z^{\prime}$ has non-vanishing
coupling to the SM-leptons, $Z^{\prime} \to l^+ l^-$ is a
promising channel for the search for $Z^{\prime}$ at the LHC.
Finally, the the muon anomalous magnetic moment was considered in
this low $B-L$ extension of the SM. $Z^{\prime}$ exchange one-loop
diagram leads to the dominant new contribution. We found that
$a_{\mu}(Z^{\prime})$ is of order $few~ \times 10^{-11}$, which
can not accommodate the $3.4 \sigma$ discrepancy between the
experimental result and the SM prediction.
\section*{Acknowledgment}
This work is partially supported by the ICTP through the
OEA-project-30. I would like to thank M. Abbas, W. Emam and R.
Mohapatra for their useful discussions.


\end{document}